\documentclass{ws-ijmpcs}

\begin{document}

\markboth{Hatsagortsyan, Kryuchkyan}
{Photon-photon interaction in structured QED vacuum}

%
\catchline{}{}{}{}{}
%

\title{PHOTON-PHOTON INTERACTION IN STRUCTURED QED VACUUM}

\author{K. Z. HATSAGORTSYAN}

\address{Max-Planck-Institut f\"ur Kernphysik\\
Saupfercheckweg 1, D-69117 Heidelberg, Germany\\
k.hatsagortsyan@mpi-k.de}

\author{G. Yu. KRYUCHKYAN}

\address{Yerevan State University, Alex Manoogian Street 1\\
Yerevan 0025, Armenia\\
kryuchkyan@ysu.am}

\maketitle

\begin{history}
\received{Day Month Year}
\revised{Day Month Year}
\end{history}

\begin{abstract}
In spatially structured strong laser fields, quantum electrodynamical vacuum behaves like a nonlinear Kerr medium with modulated third-order susceptibility where new coherent nonlinear effects arise due to modulation. We consider the enhancement of vacuum polarization and magnetization via coherent spatial vacuum effects in the photon-photon interaction process during scattering of a probe laser beam on parallel focused laser beams. Both processes of elastic and inelastic four wave-mixing in structured QED vacuum accompanied with Bragg interference are investigated. The phase-matching conditions and coherent effects in the presence of Bragg grating are analyzed for photon-photon scattering.

\keywords{vacuum polarization; super-strong fields; photon-photon scattering; Bragg scattering.}
\end{abstract}

\ccode{PACS numbers: 11.25.Hf, 123.1K}

\section{Introduction}

The existence of virtual electron-positron pair fluctuations in quantum electrodynamical (QED) vacuum allows  its description as a material medium with dielectric properties. The virtual electron-positron pair fluctuations give rise to polarization and magnetization currents which depend on the intensity of the external field  similar to nonlinear optics. Hence, the QED vacuum is nonlinear in applied external electromagnetic fields, however, the nonlinearity is weak even in super-strong fields available in labs. An important property of this conception is the photon-photon interaction which is predicted to happen in QED vacuum in external super-strong electromagnetic fields in analogy with a Kerr-like optical medium \cite{1}. Direct observation of photon-photon scattering is of great scientific importance and several suggestions for the measurement of this phenomenon have been put forward up to now.  The attempts to observe photon-photon scattering with strong laser fields \cite{2} were only able to determine the upper limit of the cross section. To enhance the cross section, a third laser beam can be applied to stimulate the photon emission at photon-photon scattering \cite{3}. The possibility of photon-photon scattering with modern strong laser beams is analyzed in \cite{4,5,6,7} including four-wave mixing process \cite{5} in analogy with nonlinear optics. However, in spite of a long-standing experimental challenge no efforts have led to actual detection of photon-photon scattering among real photons. Recently, strong field laser technique is advancing rapidly, in particular, aimed at the laser fusion program \cite{8}. The Extreme Light Infrastructure (ELI) \cite{9} is under development which will provide unprecedented  strong laser fields with intensities reaching up to $I=10^{26}$ W/cm$^{2}$.  These techniques are opening a door for new theoretical and experimental investigations of vacuum polarization effects \cite{10,11,Salamin}. It is expected that the next generation high-power laser systems will enable intensity regimes where the photon-photon interaction may be directly observable. Usually strong fields are produced with a tight focusing of a laser beam. In this respect, the strong field amplitudes are essentially nonuniform. Several recent works analyze the consequences of the vacuum nonuniform polarization.  Among these we note the investigation of diffraction and interference effects due to the vacuum nonuniform polarization in strong focused laser beams \cite{12,13}.  In analogy with nonlinear optics, it is expected that coherent spatial effects in QED vacuum could be also realized. In this regard, a new phenomenon -- Bragg scattering of light in vacuum structured by strong periodic fields (magnetic or laser) have been proposed in \cite{14,15}.  It has been shown that Bragg scattering setup provides an enhancement of the number of scattered photons with respect to the stimulated photon-photon scattering by a factor proportional to the square of the number of grating periods within the interaction region.

In this paper, we continue investigation of coherent spatial vacuum effects in photon-photon interaction in a setup of multiple crossed superstrong laser beams. The previous consideration \cite{15} was devoted to investigation of  elastic photon-photon Bragg scattering. Generally, it is difficult to distinguish QED scattered photons with wavelengths  close to that of the incoming laser beams in standard schemes. However, in our scheme of elastic photon-photon scattering the Bragg effect facilitates the observation because the Bragg interference arising at a specific impinging direction of the probe wave concentrates the scattered light in the specular direction. Here we propose an alternative  approach employing general properties of four-wave mixing processes in structured QED vacuum. In addition to elastic process, we analyze possibilities for observation of vacuum polarization effects using inelastic photon-photon scattering processes.

The paper is organized as follows: In Sec. II, we introduce the nonlinear equations accounting for the cubic-order QED vacuum polarization effects. In Sec. III, we calculate the intensity of output light in four-wave mixing processes if the incoming wave vectors satisfy the phase-matching conditions.  In Sec. IV, we discuss the phase-matching condition and coherent effects in inelastic photon-photon scattering.

\section{Nonlinear current induced by vacuum polarization}

\begin{figure}
\begin{center}
\includegraphics[width=0.3\textwidth]{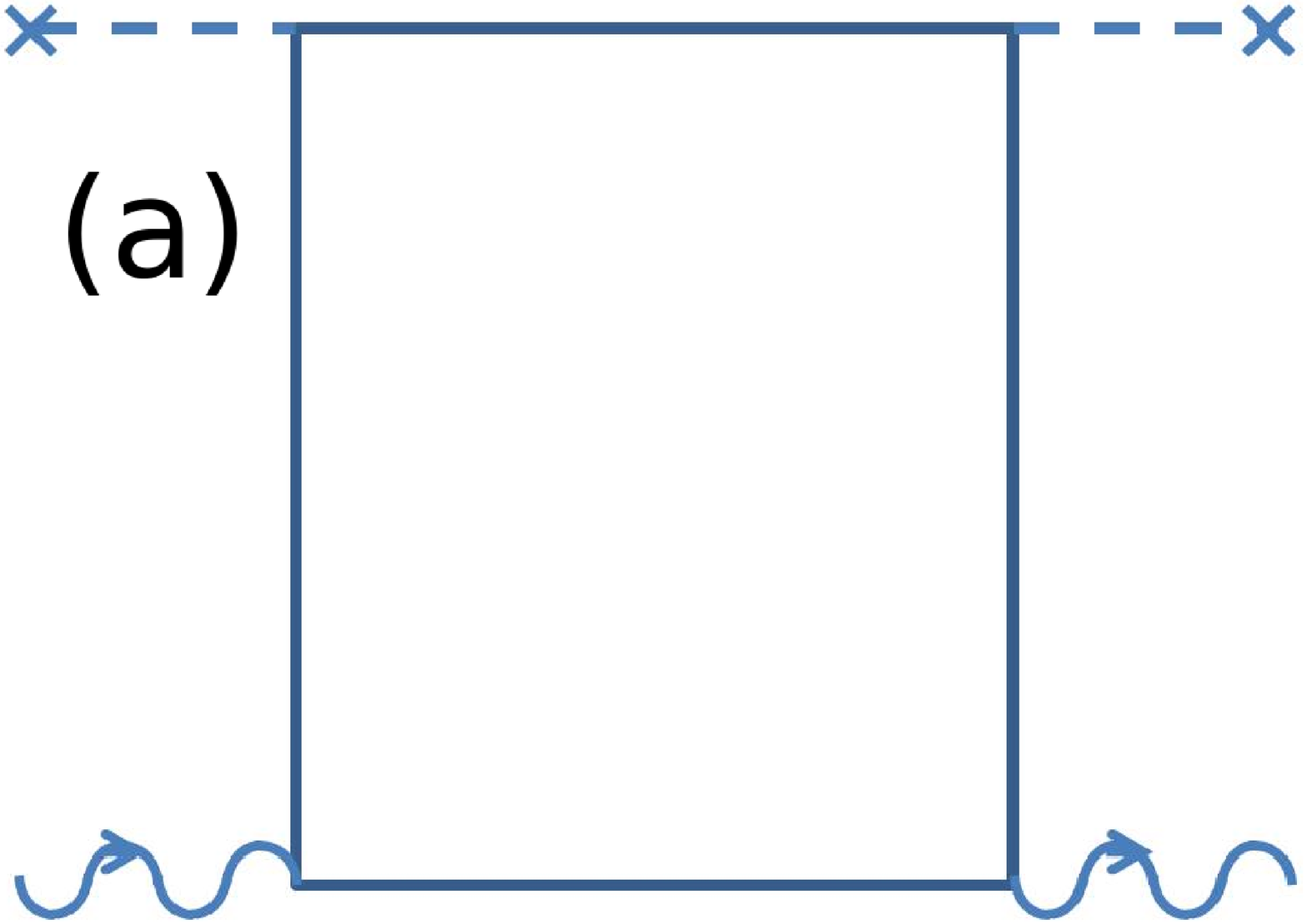}
\hspace{1cm}
\includegraphics[width=0.45\textwidth]{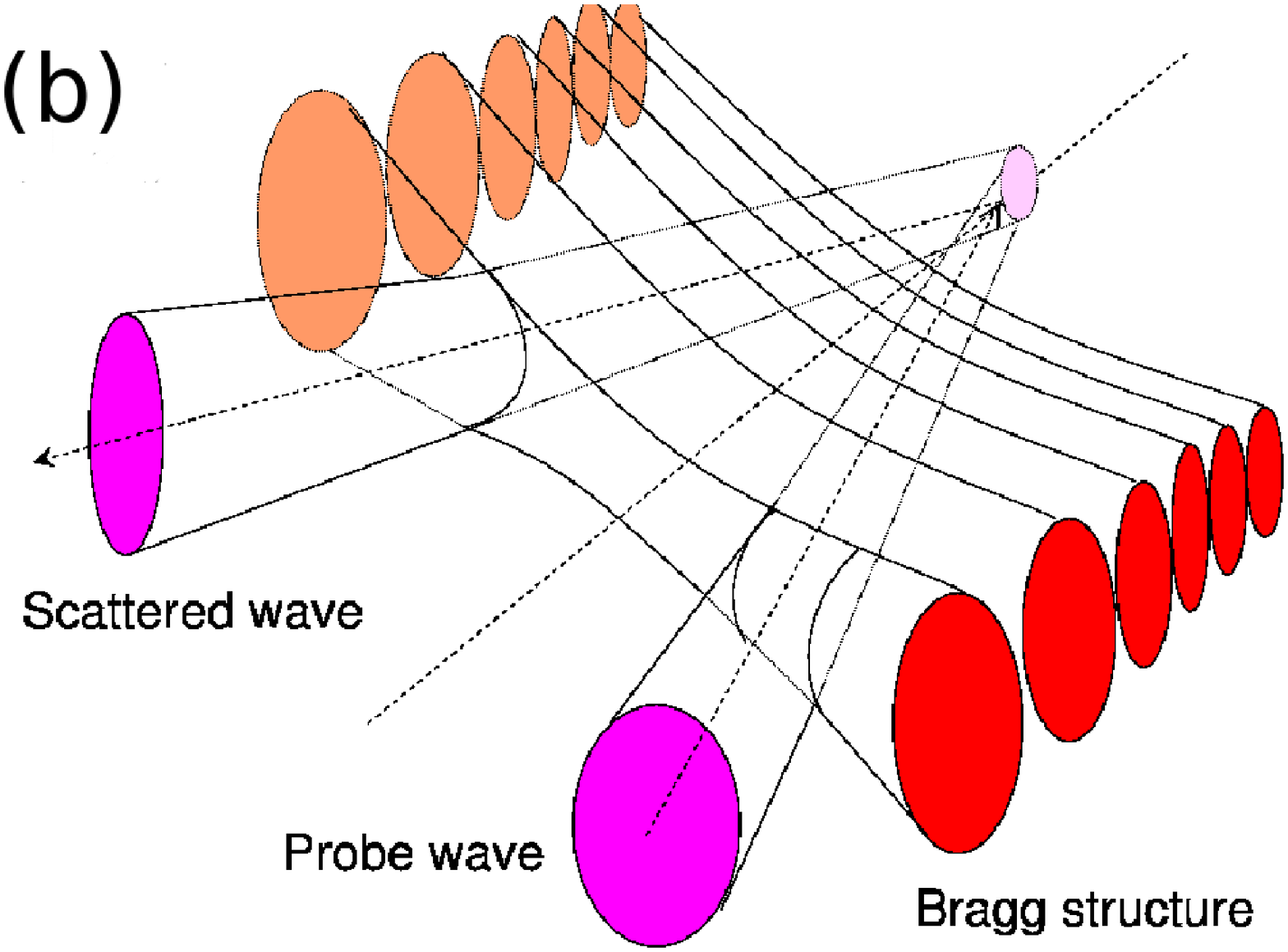}
\end{center}
\caption{  (a) The Feynman diagram describing the photon scattering in a strong external electromagnetic field. (b) Bragg scattering of a probe laser beam by a set of focused strong laser beams. } \label{fig }
\end{figure}

We consider scattering of probe light beam by a periodic spatial Bragg structure that is formed by a set of N focused laser beams propagating parallel to each other, see, Fig. 1 (b). The photon-photon interaction is described in QED by the box Feynman diagram, see Fig. 1 (a), according to which the interaction of two incoming photons produces two new outgoing photons, with photon energies and momenta fulfilling the conservation laws. One of the incoming photons belongs to a probe laser beam and the second one to the modulated structure of strong parallel laser beams. The scattered photon comprises the one of outgoing waves and the strong laser wave contributing to the second outgoing wave. 
Thus, this process can be also considered as a four wave-mixing in which interaction of three beams leads to generation of output mode satisfying resonance conditions between the frequencies and wave-vectors. In the presence of the amplitude modulation of the strong laser field, the  probe beam can be reflected by the periodic structure due to Bragg scattering which corresponds to the elastic scattering of the photons of the probe beam \cite{15}. In the following, we consider the general case of the  Bragg scattering of probe photons which can be as elastic (one laser photon is absorbed and one laser photon emitted) as well as inelastic (two laser photons are absorbed or emitted).  
The photon-photon interaction is a perturbation and we assume that it does not modify significantly the dispersion relation for laser and probe waves.

We start from the Maxwell wave equations for the electric field in the presence of the vacuum polarization $\textbf{P}$ and magnetization $\textbf{M}$
\begin{equation}
\nabla^{2}\textbf{E}-\frac{\partial^{2}\textbf{E}}{\partial t^{2}}=4\pi\frac{\partial\textbf{j}}{\partial t}+\nabla(\nabla\textbf{E}),
\end{equation}
where the current density $\textbf{j}$ is determined by the vacuum polarization $\textbf{P}$ and magnetization $\textbf{M}$:
\begin{equation}
\textbf{j}=\frac{\partial\textbf{P}}{\partial t}+\nabla\times\textbf{M}.
\end{equation}
The vacuum polarization and magnetization in the third order nonlinear corrections are derived from the Heisenberg– Euler effective Lagrangian \cite{1}:
\begin{equation}
\textbf{P}(\textbf{r},t)=\frac{\eta}{4\pi}\left[2(\textbf{E}^{2}-\textbf{B}^{2})\textbf{E}+7(\textbf{E}\cdot\textbf{B})\textbf{B}\right],
\end{equation}
\begin{equation}
\textbf{M}(\textbf{r},t)=\frac{\eta}{4\pi}\left[-2(\textbf{E}^{2}-\textbf{B}^{2})\textbf{B}+7(\textbf{E}\cdot\textbf{B})\textbf{E}\right],
\end{equation}
where $\hbar=c=1$ units are used throughout the paper.
Bellow, we derive the nonlinear wave equation for the output field (scattered light in Fig. 1), considering the current in the first-order of the probe plane wave $\textbf{E}_{in}$ and in the second-order of the composite field $\textbf{E}_{L}$  which describes  the set of parallel laser beams.  The coupling constants in the current density may be interpreted in terms of the nonlinear susceptibility of the vacuum. Choosing the total electric and magnetic fields $\textbf{E}$ and $\textbf{B}$ in  Eqs. (3) and (4) as a superposition of the composite  and the probe  field,  we obtain in this approximation
\begin{equation}
\textbf{P}^{(1)}(\textbf{r},t)=\frac{\eta}{4\pi}\left\{4\left[(\textbf{E}_{in}\textbf{E}_{L})-(\textbf{B}_{in}\textbf{B}_{L})\right]\textbf{E}_{L}+7\left[(\textbf{E}_{in}\textbf{B}_{L})+(\textbf{B}_{in}\textbf{E}_{L})\right]\textbf{B}_{L}\right\},
\end{equation}
\begin{equation}
\textbf{M}^{(1)}(\textbf{r},t)=\frac{\eta}{4\pi}\{-4[(\textbf{E}_{in}\textbf{E}_{L})-(\textbf{B}_{in}\textbf{B}_{L})]\textbf{B}_{L}+7[(\textbf{E}_{in}\textbf{B}_{L})+(\textbf{B}_{in}\textbf{E}_{L})]\textbf{E}_{L}\},
\end{equation}
where the superscript (1) denotes the terms containing the probe field in the first order. The incoming wave is chosen to be a plane wave with an amplitude  having a weak space-time dependence due to interaction 
\begin{equation}
\textbf{E}_{in}=\textbf{E}_{1}(r,t)e^{-i(\omega_{1}t-\textbf{k}_{1}\textbf{r})}+\rm c.c.\,\,\,.
\end{equation}
In accordance to the discussion above, the square of the driving composite laser fields can be written as  
\begin{equation}
\textbf{E}_{L}^{2}=\textbf{E}_{0}^{2}F(x,y,z)\cos^{2}(\omega_{L}t-\textbf{k}_{L}\textbf{r}+\varphi),
\end{equation}
where $F(x,y,z)$ describes the spatial modulation of the intensity and $\varphi$ the phase modulation ($E_{L}^{2}=B_{L}^{2}$). Using  Eqs.(2)-(8),  the current is calculated
\begin{equation}
\textbf{j}(r,t)=\textbf{j}^{(+)}(r,t)+\textbf{j}^{(-)}(r,t),
\end{equation}
where $\textbf{j}^{(-)}(r,t)=(\textbf{j}^{(+)}(r,t))^{*}$,  and
\begin{equation}
\textbf{j}^{(+)}(r,t)=-i\omega_{1}\frac{\eta}{4\pi}\cos^{2}(\omega_{L}t-\textbf{k}_{L}\textbf{r}+\varphi)F(x,y,z)e^{-i(\textbf{k}_{1}\textbf{r}-\omega_{1}t)}\textbf{S},
\end{equation}
with $\eta=e^4/(45\pi m^4)$, the electron charge and mass $e$ and $m$, respectively, and
\begin{eqnarray}
\textbf{S}=4[{(\textbf{E}_{1}\textbf{E}_{0}})-((\hat{k}_{1}\times\textbf{E}_{1})\textbf{B}_{0})]\bigg[\textbf{E}_{0}+\bigg(\bigg(\hat{k}_{1}-i\frac{\boldsymbol{\nabla} F}{k_{1}F}\bigg)\times\textbf{B}_{0}\bigg)\bigg]+\nonumber\\
7[{(\textbf{E}_{1}\textbf{B}_{0}})+((\hat{k}_{1}\times\textbf{E}_{1})\textbf{E}_{0})]\bigg[\textbf{B}_{0}-\bigg(\bigg(\hat{k}_{1}-i\frac{\boldsymbol{\nabla} F}{k_{1}F}\bigg)\times\textbf{E}_{0}\bigg)\bigg].
\end{eqnarray}
Only the resonant terms including the factors $E_{1}E_{0}^{2}$ and $B_{1}B_{0}^{2}$ are kept in the current. All other terms will be cancelled when averaging over a short space-time interval due to rapid oscillations.

\section{Four-wave mixing in structured vacuum }

In this section we calculate the intensity of light generated by the resonant four-wave interaction. For this purpose, we consider the wave equation (1) for the output field $\textbf{E}_{out}=\textbf{E}_{2}\exp(-i\omega_{2}t) +c.c.$ with a cubic nonlinear current of Eq. (10). The solution to the wave equation  in the radiation zone takes the form of a spherical outgoing wave multiplied by a phase-matching function as well as by the photon polarization matrix element.  The phase-matching function is peaked at the resonant values of the wave vector $\textbf{k}_{2}$. We present the final result for elastic photon scattering when frequencies of the probe and the output fields are the same $(\omega_{2}=\omega_{1})$, as well as the result for inelastic scattering $(\omega_{2}=\omega_{1}\pm 2\omega_{L})$ which takes place with absorption or emission of additional  photons $\omega_{L}$ from the set of parallel lasers. We also average the current over the fast oscillations of nonresonant terms which do not fulfill the phase-matching condition. For the elastic scattering this procedure leads to the replacement $\cos^{2}(\omega_{L}t-\textbf{k}_{L}\textbf{r}+\varphi)\rightarrow\frac{1}{2}$ in Eq.~(10), yielding for the electric field  
\begin{equation}
\textbf{E}_{2}(r)=\frac{1}{R_{0}}e^{-i\omega_{2}t}\int\bigg(\frac{\partial}{\partial t}\textbf{j}^{(+)}(r',t)\bigg)_{t-R_{0}+\textbf{r}'\hat{n}}dV',
\end{equation}
where the integration is carried out over the interaction region, and the current  is  evaluated at the retarded time $t=t-|\textbf{R}_{0}-\textbf{r}'|/c$,  $\hat{\textbf{n}}$ is the unit vector in the direction of $\textbf{k}_{2}$ and $\textbf{R}_{0}$ is the radius-vector of the observation point. Thus, in the case of elastic scattering, we obtain
\begin{equation}
\textbf{E}_{2}(r)=-\omega_{1}^{2}\frac{\eta}{8\pi}\frac{e^{ik_{2}R_{0}}}{ R_{0}}\textbf{S}\int F(x',y',z')e^{i(\textbf{k}_{1}-\textbf{k}_{2})\textbf{r}'}dV'.
\end{equation}
In the case of inelastic scattering, the factor of $\cos^{2}(\omega_{L}t-\textbf{k}_{L}\textbf{r}+\varphi)$   modifies the resonant terms and the radiated  field is calculated as follows
\begin{equation}
\textbf{E}_{2}(r)=\omega_{1}(2\omega_{L}\pm\omega_{1})\frac{\eta}{8\pi}\frac{e^{ik_{2}R_{0}}}{R_{0}}\textbf{S}\int F(x',y',z')e^{i(\textbf{k}_{1}\pm 2\textbf{k}_{L}-\textbf{k}_{2})\textbf{r}'}dV'.
\end{equation}
It corresponds to two configurations with absorption or emission of laser photons which fulfill the energy conservation: 
\begin{equation}
\omega_{2}=\omega_{1}\pm 2\omega_{L}.
\label{omega}
\end{equation}

We calculate the intensity of the scattered field within a solid angle $d\Omega$: $dI=(1/4\pi)R_0^2|\sum_{\lambda} E_{2\lambda}(r)|^{2}d\Omega$,  summing over polarization of the scattered photons $\lambda$.
The intensity of generated mode during elastic photon scattering reads
\begin{equation}
\frac{dI_{el}}{d\Omega}=\frac{\eta^{2}\omega_{1}^{4}}{32(2\pi)^{3}}|S|^{2}P(\Delta\textbf{k}_{1}),
\label{Iel}
\end{equation}
while  for the inelastic case reads as
\begin{equation}
\frac{dI_{inel}}{d\Omega}=\frac{\eta^{2}(2\omega_{L}-\omega_{1})^{2}\omega_{1}^{2}}{32(2\pi)^{3}}|S|^{2}P(\Delta\textbf{k}_{2}),\label{Iinel}
\end{equation}
\begin{equation}
P(\Delta\textbf{k})=\bigg|\int F(r') e^{i\Delta\textbf{k}\textbf{r}'}dV'\bigg|^{2},
\end{equation}
where  $\Delta\textbf{k}=\Delta\textbf{k}_{el}\equiv\textbf{k}_{1}-\textbf{k}_{2}$ is the phase-mismatch function in the case of elastic scattering and $\Delta\textbf{k}=\Delta\textbf{k}_{inel}\equiv\textbf{k}_{1}\pm 2\textbf{k}_{L}-\textbf{k}_{2}$ in the case of inelastic scattering.
In Eqs. (\ref{Iel}) and (\ref{Iinel}), $|S|^{2}$  depends from polarization states of the probe and of the driving fields.
We see that the efficiency of the scattering is determined by the Fourier transformation of the intensity distribution of the driving strong laser field which forms Bragg grating. The distribution function $F(x,y,z)$ for $N$ parallel laser beams reads
\begin{equation}
F(x,y,z)=\sum_{n=1}^{N}f(x,y,z-nd),
\end{equation}
where $d$ is the period of the Bragg structure and $f(x,y,z)$ is the  distribution function of the single laser beam. The latter is chosen in a form  of an elliptical Gaussian beam 
\begin{equation}
f(x,y,z)=e^{\frac{2y^{2}}{w_{y}^{2}}-\frac{2z^{2}}{w_z^{2}}},
\end{equation}
where $w_{y,z}$ are the beam waist sizes. After the integration over the interaction region, we have for the phase-mismatch function
\begin{equation}
P(\Delta\textbf{k})=\frac{\pi^2}{2}L_x(w_yw_z)^2\exp\left(-\frac{\Delta k_y^2w_y^2}{4}-\frac{\Delta k_z^2w_z^2}{4}\right)\left|\sum_n^N\exp\left(i\Delta k_znd\right)\right|^2\delta (\Delta k_x),
\end{equation}
where $L_x$ is the interaction length in the $x$-direction and $\delta (x)$ is the Dirac delta-function. Taking into account that
\begin{equation}
\left|\sum_n^N\exp\left(i\Delta k_znd\right)\right|^2=\frac{\sin^2\frac{\Delta k_zNd}{2}}{\sin^2\frac{\Delta k_zd}{2}}=\frac{2\pi}{d}N\sum_l \delta \left(\Delta k_z-\frac{2\pi l}{d}\right),
\end{equation}
the phase-mismatch function reads
\begin{equation}
\label{PP}
P(\Delta\textbf{k})=\frac{\pi^3}{d}N L_x(w_yw_z)^2\exp\left(-\frac{\Delta k_y^2w_y^2}{4}-\frac{\Delta k_z^2w_z^2}{4}\right) \delta (\Delta k_x)\sum_l \delta \left(\Delta k_z-lq_z\right),
\end{equation}
where $\textbf{q}=(2\pi/d )\hat{\textbf{z}}$, $\hat{\textbf{z}}$ is the unit vector along the $z$-axis and $l$ is an integer number. The phase-mismatch function $P(\Delta\textbf{k})$ generates the phase-matching Bragg condition 
\begin{equation}
\Delta\textbf{k}-l\textbf{q}=0.
\label{bragg}
\end{equation}
The exponential damping factor in Eq. (\ref{PP}) determines the maximal value for $\Delta k_z\lesssim  2/w_z$. Taking into account the Bragg condition Eq. (\ref{bragg}), the damping factor restricts the harmonic number  $l\lesssim  d/\pi w_z$.

The total intensity of the scattered radiation, integrated over the solid angle, equals:
\begin{eqnarray}
I_{el}&=&\frac{\eta^2\omega_1^2}{128(2\pi)^{3/2}}|S|^2\left( L_xw_yw_z\tau\right)N, \\
I_{inel}&=&\frac{\eta^2(2\omega_L-\omega_1)^2}{128(2\pi)^{3/2}}|S|^2\left( L_xw_yw_z\tau\right)N,
\label{total}
\end{eqnarray}
where $\tau$ is the interaction time. The total intensity is proportional to the interaction 4-volume $L_xw_yw_z\tau$ as well as to the number of periods in the Bragg structure $N$.
Thus, the lattice structure of a medium can significantly ($\sim N$) enhance the intensity of the scattered light due to  interference of the scattered light generated from different layers of the structure (Bragg scattering) [16].

Note that the Bragg concept is quite general and is applied not only in optics [17] but also in quantum optics [18], atom optics [19] and for matter waves [20].

\section{Phase-matching and coherent effects }

\begin{figure}
\begin{center}
\includegraphics[width=0.5\textwidth]{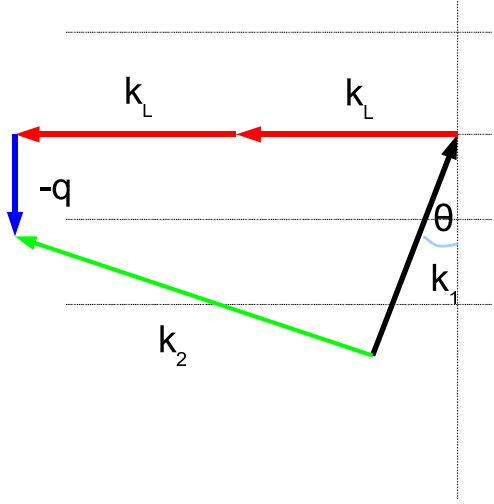}
\end{center}
\caption{To the phase-matching condition  in the case of inelastic four-wave mixing.} 
\label{fig2}
\end{figure}

In this section, we analyze the phase-matching condition of Eq. (\ref{bragg}) in the case of inelastic four-wave mixing, with the energy conservation given by Eq. (\ref{omega}). Let us express both of the above equations via 4-vectors: 
\begin{equation}
k_2=k_1-q\pm 2k_L,
\label{4-vector}
\end{equation}
with $q=(0,0,0,q_0)$, $q_0\equiv 2\pi/d$, and $k_{1,2,L}=(\omega_{1,2,L},\textbf{k}_{1,2,L})$, and find the phase-matching condition for the impinging angle of the probe wave, see Fig. 2. Taking into account that $k_L\cdot q=0$, we find from the squaring of Eq. (\ref{4-vector}):
\begin{equation}
-q_0^2+2\omega_1q_0 \cos \vartheta \pm 4\omega_1\omega_L(1-\sin\vartheta)=0.
\label{4-vector2}
\end{equation}
The latter yields to a quadratic equation for $\tan \frac{\vartheta}{2}$:
\begin{equation}
 \left(\tan \frac{\vartheta}{2}\right)^2(q_0^2+2\omega_1q_0)\mp 8\omega_1\omega_L\left(\tan \frac{\vartheta}{2}\right)+(q_0^2-2\omega_1q_0)=0.
\label{4-vector3}
\end{equation}
This equation will have a solution when
\begin{equation}
 16\omega_1^2\omega_L^2+4\omega_1^2q_0^2>q_0^4,
\label{D}
\end{equation}
and it reads
\begin{equation}
\vartheta=\tan^{-1}\left\{\frac{\pm 4\omega_1\omega\pm \sqrt{16\omega_1^2\omega_L^2+4\omega_1^2q_0^2}}{q_0^2+2\omega_1q_0} \right\}.
\label{tg}
\end{equation}
Usually, the distance between the laser beams in the Bragg structure should be larger than the waist size of the beam: $d>w_z>\lambda_L$, from which it follows that $q_0<\omega_L$. For example, considering $q_0=\omega_L/2$ and $\omega_1=\omega_L$ [in which case $\omega_2=3\omega_L$], the condition of  Eq. (\ref{D}) is fulfilled, i.e. the phase-matching is possible to fullfill, and Eq. (\ref{tg}) yields for the probe impinging angle $\vartheta \approx -8.5^o$, see Fig. 2. In the chosen geometry, the scattered wave has larger frequency and emission direction significantly varying from the probe direction.

\section{Conclusion}

We have studied scattering of probe laser beam on the set of parallel focused laser that constitutes a Bragg grating. Both processes of elastic and inelastic four wave-mixing in structured QED vacuum accompanied with  Bragg interference are investigated.  In the case of elastic scattering of photons, when the frequencies of the probe wave $\omega_{1}$ and the scattered wave $\omega_{2}$ are the same, the phase-matching condition is met  at $\textbf{k}_{1}=\textbf{k}_{2}+\textbf{q}$, where $\textbf{q}$ is the vector of the lattice. While, for inelastic light scattering with absorption or emission of additional laser photons i.e. for the processes $\omega_{1}=\omega_{2}+2\omega_{L}$, or $\omega_{1}=\omega_{2}-2\omega_{L}$, the phase-matching condition is accordingly modified $\textbf{k}_{1}=\textbf{k}_{2}\pm2\textbf{k}_{L}+\textbf{q}$. 
When the phase-matching condition is fulfilled, the scattering probability is enhanced  due to Bragg interference over the incoherent one by a factor $N$ of the number of periods in the grating. While in the elastic scattering process, the scattered wave can be distinguished from the probe only by the emission direction, in the inelastic process, it is distinguished both by the frequency and by the emission direction. This significantly improves the experimental feasibility for the observation of vacuum polarization effects with strong laser fields.

\end{document}